\documentclass[aps,prd, twocolumn,superscriptaddress,nofootinbib,preprintnumbers]{revtex4-1}
\usepackage{amsmath}
\usepackage{graphicx}
\usepackage{subfigure}
\usepackage{amssymb}
\usepackage{xcolor}
\usepackage{multirow}
\usepackage{cancel}
\usepackage{color}
\usepackage{ulem}
\usepackage{listings}
\usepackage{xcolor}
\usepackage{float}
\usepackage{slashed}

\usepackage[colorlinks,citecolor=blue]{hyperref}
\usepackage{amsmath}
\usepackage{wrapfig}

\usepackage[utf8]{inputenc} 

\newcommand{\be}{\begin{equation}}
\newcommand{\ee}{\end{equation}}
\newcommand{\bea}{\begin{eqnarray}}
\newcommand{\eea}{\end{eqnarray}}
\newcommand{\besp}{\begin{equation}\begin{split}}
\newcommand{\eesp}{\end{split}\end{equation}}
\newcommand{\met}{\not{\!\!{\rm E}}_{T}}
\newcommand{\nn}{\nonumber}

\newcommand{\Eq}[1]{Eq.~(\ref{#1})}

\newcommand{\Dfbd}{\mathord{\buildrel{\lower3pt\hbox{$\scriptscriptstyle\leftrightarrow$}}\over {D}_{\mu}}}

\def\mL{\mathcal{L}}

\def\mN{\mathcal{N}}
\def\mO{\mathcal{O}}

\def\1{\textbf{1}}
\def\2{\textbf{2}}
\def\3{\textbf{3}}
\def\4{\textbf{4}}
\def\5{\textbf{5}}
\def\6{\textbf{6}}
\def\7{\textbf{7}}
\def\8{\textbf{8}}
\def\9{\textbf{9}}

\begin{document}

\title{Beyond $M_{t\bar{t}}$: learning to search for a broad $t\bar t$ resonance at the LHC}

\author{Sunghoon Jung}
\email{sunghoonj@snu.ac.kr}
\affiliation{Center for Theoretical Physics, Department of Physics and Astronomy, Seoul National University, Seoul 08826, Korea}

\author{Dongsub Lee}
\email{dongsub93@snu.ac.kr}
\affiliation{Center for Theoretical Physics, Department of Physics and Astronomy, Seoul National University, Seoul 08826, Korea}

\author{Ke-Pan Xie}
\email{kpxie@snu.ac.kr}
\affiliation{Center for Theoretical Physics, Department of Physics and Astronomy, Seoul National University, Seoul 08826, Korea}

\begin{abstract}

A resonance peak in the invariant mass spectrum has been the main feature of a particle at collider experiments. However, broad resonances not exhibiting such a sharp peak are generically predicted in new physics models beyond the Standard Model. Without a peak, how do we discover a broad resonance at colliders? We use machine learning technique to explore answers beyond common knowledge. We learn that, by applying deep neural network to the case of a $t\bar{t}$ resonance, the invariant mass $M_{t\bar{t}}$ is still useful, but additional information from off-resonance region, angular correlations, $p_T$, and top jet mass are also significantly important. As a result, the improved LHC sensitivities do not depend strongly on the width. The results may also imply that the additional information can be used to improve narrow-resonance searches too. Further, we also detail how we assess machine-learned information.

\end{abstract}

\maketitle

\newpage
\section{Introduction}

Discovering new physics through a new resonance is one of the most exciting opportunities. A ``narrow'' resonance peak, being sharply localized in the energy spectrum, allows for the most efficient discovery above continuum backgrounds as well as for precision measurements of the particle mass, width and other properties. However, the widths of new (and presumably heavier) resonances in new physics can be easily much larger than those of the Standard Model (SM) particles. The width generally grows with the mass of a resonance, and a new strong coupling may induce rapid decay as in composite Higgs models~\cite{Barducci:2012kk,Greco:2014aza,Barducci:2015vyf,Liu:2019bua} or warped extra dimensional models~\cite{Kelley:2010ap,Ask:2011zs}. Also, more decay channels to lighter beyond-SM particles may open up, which further increases the width.

The large width causes several difficulties in collider experiments. Above all, without a sharp peak, the discovery becomes challenging, as the signal becomes spread over a large range of energy above continuum backgrounds. For example, the ATLAS result based mostly on the invariant mass distribution~\cite{Aaboud:2018mjh} shows that for a $M=1$ TeV Kaluza-Klein gluon, the measured (expected) cross section upper limit $\sigma(pp\to g_{KK}\to t\bar t)$ increases from 1.4 (1.2) pb to 4.7 (2.7) pb when the width-to-mass ratio $\Gamma/M$ varies from 10\% to 40\%. In addition, the phenomenological study in Ref.~\cite{Liu:2019bua} shows that for the minimal composite Higgs model with the third generation left-handed quark $q_L=(t_L,b_L)^T$ being fully composite, a vector $t\bar t$ resonance as light as $M=1$ TeV is still allowed by the direct search in the $\Gamma/M\gtrsim20\%$ region. 

Secondly, broad resonance shape is more susceptible to the energy dependences of parton luminosity and the width, interferences with backgrounds or other resonances, and mixing and overlap with nearby resonances. These effects make discoveries further challenging and complicated. In particular, the complex interference (the one with imaginary parts in amplitudes) in supersymmetric or two-Higgs doublet models can make broad heavy Higgs bosons decaying to $t\bar{t}$ generally appear not as a pure resonance peak~\cite{Gaemers:1984sj,Dicus:1994bm,Craig:2015jba,Jung:2015gta} but even as pure dips or nothing~\cite{Jung:2015gta}. And nearly degenerate heavy Higgs bosons can overlap significantly, producing complicated resonance shapes~\cite{Choi:2004kq,Ellis:2004fs,Carena:2016npr}. 

Many of these new broad resonances are just beyond the current reach of the LHC. Thus, it is imperative to study the physics of broad resonances and develop efficient discovery methods. However, broad-resonance searches have been studied only in limited cases, e.g., phenomenologically in Refs.~\cite{Liu:2019bua} (third-generation quark pair, $\ell^+\ell^-$),~\cite{Kelley:2010ap} ($\mu^+\mu^-$), and experimentally in Refs.~\cite{Aaboud:2018mjh,Sirunyan:2018ryr} ($t\bar t$),~\cite{Sirunyan:2018xlo,Aaboud:2016nbq} ($jj$)~\cite{Aaboud:2017buh}($\ell^+\ell^-$). In all the cases, the invariant mass had still been used as a main observable, but the question of ``how do we (best) discover a broad resonance without a peak?'' had not been answered thoroughly~\footnote{If a broad resonance can decay to multi-top/$W$ channels, it can be searched using the same-sign di-lepton final state~\cite{Barducci:2015vyf,Liu:2019bua,Liu:2018hum}, which doesn't rely on the reconstruction of the invariant mass.}.

This question might be a problem appropriate to use deep neural network (DNN) technique to answer. It is because the answer is not so obvious, {\it a priori}, and even small improvements will be significant. Machine learning has indeed been applied to various problems in particle physics. For example, bump-hunting resonance searches were improved with DNN~\cite{Collins:2018epr,Collins:2019jip}. The DNN is one of machine learning algorithms. Coming with various network structures such as fully-connected network~\cite{Hajer:2018kqm,Baldi:2014kfa,Baldi:2016fzo,Luo:2017ncs,Lee:2018xtt,Pearkes:2017hku}, convolutional neural network~\cite{Guo:2018hbv,deOliveira:2015xxd,Cogan:2014oua,Li:2019ufu} and others~\cite{Fraser:2018ieu,Louppe:2017ipp,Abdughani:2018wrw,Ren:2019xhp,henrion2017neural}, DNN had shown remarkable performances in the exploration of physics beyond the SM, often better than other machine learning algorithms such as boosted decision tree (BDT). We refer to Refs.~\cite{Guest:2018yhq,Abdughani:2019wuv} and references therein for reviews of the DNN applications in LHC physics. 

In this paper, we consider a spin-1 broad $t\bar{t}$ resonance at the LHC (Sec.~\ref{sec:benchmark}). Being the heaviest particle in the SM, the top quark has been regarded as an important portal to new physics. As a first step toward a more general study of broad resonances, we ignore any interference effects and nearby resonances (Sec.~\ref{sec:DNN}). We use fully-connected DNN to explore answers beyond common knowledge (Sec.~\ref{sec:DNN}).  Finally, we assess whether and what DNN can learn, even beyond what we know well (Sec.~\ref{sec:learn}).

\section{Benchmark model}\label{sec:benchmark}
 
For simplicity, here we consider a gauge singlet vector resonance $\rho$ interacting strongly with the SM right-handed top quark $t_R$, and the relevant Lagrangian is
\bea\label{eq:Lagrangian}
\mL=&&~-\frac{1}{4}\rho_{\mu\nu}\rho^{\mu\nu}+\frac{m_{\rho}^2}{2 g_{\rho}^2}(g_{\rho}\rho_{\mu}-g_1B_\mu)^2\nn \\ 
&&~+\bar t_R\gamma^\mu t_R(g_{\rho}\rho_{\mu}-g_1B_\mu),
\eea
where $\rho_{\mu\nu}=\partial_\mu\rho_{\nu}-\partial_\nu\rho_{\mu}$, and $g_1$ is the SM hypercharge gauge coupling. This model is also considered in Refs.~\cite{Greco:2014aza,Liu:2018hum,Liu:2015hxi}. Note that the $\rho_\mu$ mixes with the SM gauge field $B_\mu$ in \Eq{eq:Lagrangian}. Given $g_\rho\gg g_1$, the mixing angle is $\sin\theta\approx g_1/g_\rho$ before the electroweak symmetry breaking (EWSB). Therefore, after transforming to the mass eigenstates, the interactions between $\rho$ resonance and SM fermions will be $\sim g_\rho$ for $t_R$ and $\sim Yg_1^2/g_\rho$ for other fermions (including $t_L$ and other light quarks), with $Y$ being the hypercharge of the corresponding fermion. The physical mass of $\rho$ is $M_\rho=m_\rho$. EWSB gives $\mO(v^2/m_\rho^2)$ corrections to above picture, and the details can be found in Appendix C of Ref.~\cite{Liu:2018hum}.

Due to the large coupling $g_\rho$, the $\rho$ resonance decays to $t\bar t$ with a branching ratio $\sim100\%$, and the width-to-mass ratio is
\bea
\frac{\Gamma_{\rho}}{M_\rho}\approx\frac{\Gamma_{\rho\to t\bar t}}{M_\rho}\approx\frac{g_\rho^2}{8\pi}.
\eea
For $g_\rho=3$ and 4, this ratio reaches 36\% and 64\%, respectively. Thus a broad $\rho$ is easily realized in the model described by Eq.~(\ref{eq:Lagrangian}). Note that $\Gamma_{\rho\to t\bar t}\leqslant\Gamma_\rho$, if $\rho$ has other strong dynamical decay channels such as the decay to low-mass top partners (which are not listed in our simplified model), typically $\Gamma_\rho$ is several times larger than $\Gamma_{\rho\to t\bar t}$, thus a large $\Gamma_\rho/M_\rho$ can be obtained even for smaller $g_\rho$. We consider $M_\rho=1$ and 5 TeV as two benchmarks, and for each mass point the width-to-mass ratios $\Gamma_\rho/M_\rho=10\%$, 20\%, 30\% and 40\% are considered. The corresponding benchmark cases are then identified as M$i\Gamma j$, with $i=1$ or 5 denoting the mass (in unit of TeV) and $j=1$, 2, 3, 4 being $10\times\Gamma_\rho/M_\rho$. For example, M1$\Gamma$4 is the benchmark for $M_\rho=1$ TeV and $\Gamma_\rho/M_\rho=40\%$.

At the LHC, the $\rho$ resonance can be produced via the Drell-Yan process ($q\bar{q} \to \rho$) through the $\rho$-light quark interaction. Among the various decay channels of the $t\bar t$, we choose to focus on the semi-leptonic final state
\bea
pp\to \rho\to t\bar t \to \ell^\pm\nu b\bar bjj.
\eea
The dominant background is then the SM $t\bar t$ process, which contributes $81\%\sim88\%$ of the total backgrounds~\cite{Aaboud:2018mjh}. For simplicity, we only consider this background. It should be emphasized that although we provide a benchmark model as physical motivation here, our results are general for all heavy singlet spin-1 resonances with top quark portal.

\section{Searching for a broad $t\bar t$ resonance}\label{sec:DNN}

In this section, we describe technical details of our work and show final cross section limits. First, we describe how we parameterize a broad resonance, and how we build learning datasets and train DNN for each benchmark signal case. Then we derive improved cross section upper limits.

\subsection{Breit-Wigner description}

We assume a single, isolated broad resonance far away from any other resonances and thresholds, and ignore any interference effects. Then we use the following Breit-Wigner description of the propagator of a broad resonance 
\be\label{eq:BW}
\frac{1}{ s-\hat{M}_\rho^2(s) + i \sqrt{s} \hat{\Gamma}_\rho(s) } \, \approx \, \frac{1}{ s-M_\rho^2 + i M_\rho\Gamma_\rho},
\ee
where the nominal resonance mass $M_\rho$ and the width $\Gamma_\rho$ are fixed constants. The energy dependence of the mass $\hat{M}_\rho(s)$ from the real part of the self-energy correction is higher-order, hence small irrespective of the large width. On the other hand, the energy dependence of the width $\hat{\Gamma}_\rho(s) \propto \sqrt{s}$ from the imaginary part can induce corrections as large as $\sim$100~(10)\% for broad resonances considered in this paper $\Gamma_\rho/M_\rho\sim 40~(20)\%$. But, within this range of the width, the resonance shape remains relatively undistorted albeit some shifts of the peak and height~\cite{Liu:2019bua,Kelley:2010ap,An:2012va}. Also, the fixed mass and width have been used in LHC searches of broad resonances~\cite{Aaboud:2018mjh,Sirunyan:2018ryr}. Thus, we use Eq.~(\ref{eq:BW}) with fixed $M_\rho$ and $\Gamma_\rho$, both for simplicity and for comparison purpose.

\subsection{Preparing training data}

The model described by \Eq{eq:Lagrangian} is written in the universal {\tt FeynRules} output file~\cite{Alloul:2013bka}. We generate parton-level events of the signals and background using 5-flavor scheme within the {\tt MadGraph5\_aMC@NLO}~\cite{Alwall:2014hca} package. All spin correlations of the final state $\ell^\pm\nu b\bar bjj$ objects are kept. The phase space integrate region is set to $|\sqrt{s}-M_\rho|\leqslant15\times\Gamma_\rho$, which is large enough for us to simulate the full on- and off-shell effect of the $\rho$ resonance.
The interference between $pp\to\rho\to t\bar t$ and the SM $t\bar t$ background is negligible~\cite{Aaboud:2018mjh}, thus not considered here. We normalize the SM $t\bar t$ cross section with the the next-to-next-to-leading order with next-to-next-to-leading logarithmic soft-gluon resummation calculation from the {\tt Top++2.0} package~\cite{Czakon:2011xx,Czakon:2013goa,Czakon:2012pz,Czakon:2012zr,Baernreuther:2012ws,Cacciari:2011hy}, and the $K$-factor is 1.63. The parton-level events are matched to $+1{\rm~jet}$ final state and then interfaced to {\tt Pythia 8}~\cite{Sjostrand:2014zea} and {\tt Delphes}~\cite{deFavereau:2013fsa} for parton shower and fast detector simulation. As for the detector setup, we mainly use the CMS configuration, but with following modifications: the isolation $\Delta R$ parameters for electron, muon and jet are set to 0.2, 0.3 and 0.5 respectively. The $b$-tagging efficiency (and mis-tag rate for $c$-jet, light-flavor jets) is corrected to $0.77$ (and $1/6$, $1/134$) according to Ref.~\cite{Aaboud:2018xpj}. We generate $\gtrsim5\times10^6$ events for the background and each signal benchmark.

\begin{table*}\scriptsize
\begin{tabular}{|c|c|c|c|c|c|c|c|c|c|c|c|}\hline
Process & Event number & Cut 1 & Cut 2 & Cut 3 & Efficiency \\
\hline
M1$\Gamma$1 & $5.00\times10^6$ & $3.32\times10^6$ & $3.02\times10^6$ & $1.81\times10^6$ & 36.3\% \\ \hline
M1$\Gamma$2 & $5.00\times10^6$ & $3.29\times10^6$ & $2.98\times10^6$ & $1.79\times10^6$ & 35.8\% \\ \hline
M1$\Gamma$3 & $3.85\times10^6$ & $2.52\times10^6$ & $2.23\times10^6$ & $1.36\times10^6$ & 35.3\% \\ \hline
M1$\Gamma$4 & $5.00\times10^6$ & $3.25\times10^6$ & $2.93\times10^6$ & $1.75\times10^6$ & 34.9\% \\ \hline\hline
SM $t\bar t$ & $4.98\times10^6$ & $2.60\times10^6$ & $2.21\times10^6$ & $1.39\times10^6$ & 28.0\% \\ \hline
\end{tabular}
\caption{The cut flows of the signals and background in resolved region. The events are generated at $1\ell^\pm+\met+{\rm jets}$ final state, where $\ell$ denotes $e$ and $\mu$. M$i\Gamma j$ denotes the benchmark case with $M_\rho=i$ TeV and $\Gamma_\rho/M_\rho=0.1\times j$.}\label{tab:cut_flow_resolved}
\end{table*}

We defined two kinematic regions. The first one is called the resolved region, in which the decay products of the top quark (i.e $\ell^\pm\nu b\bar bjj$) are identified as individual objects. This region is defined as follows

\begin{enumerate}
\item Exactly one charged lepton $\ell^\pm=e^\pm$ or $\mu^\pm$ with $p_T^{\ell}>30$ GeV and $|\eta^\ell|<2.5$. Events containing a second lepton with $p_T^\ell>25$ GeV are vetoed. 
\item $\slashed{E}_T>20$ GeV and $\slashed{E}_T+M_T^W>60$ GeV, where the $W$-transverse mass is defined as
\begin{equation*}
M_T^W=\sqrt{2p_T^\ell\slashed{E}_T\left[1-\cos\Delta\phi(p_T^\ell,\slashed{E}_T)\right]}.
\end{equation*}
\item At least four jets with $p_T^j>25$ GeV and $|\eta^j|<2.5$, and at least one of the leading four jets is $b$-tagged.
\end{enumerate}

\noindent The cuts are mainly based on Ref.~\cite{Aaboud:2018mjh}, but with some simplifications. The cut flows of the signals and backgrounds are listed in Table~\ref{tab:cut_flow_resolved}. We only consider the $M_\rho=1$ TeV benchmark cases in this kinematic region. The SM $t\bar t$ cross section is 68.9 pb taken into account the $K$-factor.

\begin{table*}\scriptsize
\begin{tabular}{|c|c|c|c|c|c|c|c|c|c|c|c|}\hline
Process & Event number & Cut 1 & Cut 2 & Cut 3 & Cut 4 & Efficiency \\
\hline
M1$\Gamma$1 & $5.00\times10^6$ & $3.32\times10^6$ & $3.02\times10^6$ & $1.17\times10^6$ & $9.61\times10^5$ & 19.2\% \\ \hline
M1$\Gamma$2 & $5.00\times10^6$ & $3.29\times10^6$ & $2.98\times10^6$ & $1.08\times10^6$ & $8.85\times10^5$ & 17.7\% \\ \hline
M1$\Gamma$3 & $5.00\times10^6$ & $3.27\times10^6$ & $2.96\times10^6$ & $1.02\times10^6$ & $8.34\times10^5$ & 16.7\% \\ \hline
M1$\Gamma$4 & $5.05\times10^6$ & $3.28\times10^6$ & $2.96\times10^6$ & $9.92\times10^5$ & $8.06\times10^5$ & 15.9\% \\ \hline\hline
M5$\Gamma$1 & $5.00\times10^6$ & $2.53\times10^6$ & $2.36\times10^6$ & $1.15\times10^6$ & $8.41\times10^5$ & 16.8\% \\ \hline
M5$\Gamma$2 & $5.00\times10^6$ & $2.72\times10^6$ & $2.52\times10^6$ & $1.19\times10^6$ & $8.76\times10^5$ & 17.5\% \\ \hline
M5$\Gamma$3 & $5.00\times10^6$ & $2.81\times10^6$ & $2.59\times10^6$ & $1.19\times10^6$ & $8.85\times10^5$ & 17.7\% \\ \hline
M5$\Gamma$4 & $5.00\times10^6$ & $2.86\times10^6$ & $2.64\times10^6$ & $1.20\times10^6$ & $8.90\times10^5$ & 17.8\% \\ \hline\hline
SM $t\bar t$ ({\tt xptj = 150}) & $1.99\times10^7$ & $1.22\times10^7$ & $1.08\times10^7$ & $1.41\times10^6$ & $1.21\times10^6$ & 6.10\% \\ \hline
\end{tabular}
\caption{The cut flows of the signals and background in boosted region. The events are generated at $1\ell^\pm+\met+{\rm jets}$ final state, where $\ell$ denotes $e$ and $\mu$. M$i\Gamma j$ denotes the benchmark case with $M_\rho=i$ TeV and $\Gamma_\rho/M_\rho=0.1\times j$. The setup {\tt xptj = 150} is to improve the event generating efficiency of the background, see the text for details.}\label{tab:cut_flow_boosted}
\end{table*}

The second kinematic region is the boosted region, in which the hadronic decay products of the top quark are combined into a fat jet. The corresponding event selection criteria is 
\begin{enumerate}
\item Exactly one charged lepton $\ell^\pm=e^\pm$ or $\mu^\pm$ with $p_T^{\ell}>30$ GeV and $|\eta^\ell|<2.5$. Events containing a second lepton with $p_T^\ell>25$ GeV are vetoed. 
\item $\slashed{E}_T>20$ GeV and $\slashed{E}_T+M_T^W>60$ GeV.
\item Exactly one top-jet with $p_T^{j_{\rm top}}>300$ GeV and $|\eta^{j_{\rm top}}|<2.0$, and satisfies $\Delta\phi(j_{\rm top},\ell^\pm)>2.3$. The top-jet is reconstructed with a $R=1.0$ cone in anti-$k_t$ algorithm, and is trimmed with $R_{\rm cut}=0.2$ and $f_{\rm cut}=0.05$~\cite{ATLAS-CONF-2015-035}. We use a simplified top-tagging procedure in event selection. The top-tagging efficiency and the mistag-rate are set to 80\% and 20\% respectively, based on Ref.~\cite{ATL-PHYS-PUB-2015-053}, which makes use of jet invariant mass and $N$-subjettiness~\cite{Thaler:2008ju,Kaplan:2008ie,Thaler:2010tr,Thaler:2011gf,Plehn:2011tg,Kasieczka:2015jma}.
\item Exactly one selected jet with $p_T^{j_{\rm sel}}>25$ GeV and $|\eta^{j_{\rm sel}}|<2.5$. In addition, the selected jet should have $\Delta R(j_{\rm sel},j_{\rm top})>1.5$ and $\Delta R(j_{\rm sel},\ell)<1.5$.
\end{enumerate}
\noindent The cuts here are again mainly based on Ref.~\cite{Aaboud:2018mjh}. and the cut flows for signals and background are listed in Table~\ref{tab:cut_flow_boosted}. In this region, we consider both $M_\rho=1$ and 5 TeV signals. To increase the event generating efficiency of the background events, in this region we require the SM $pp\to t\bar t\to \ell^\pm\nu b\bar bjj$ process has at least one final state parton (including the $b$-parton) with $p_T>150$ GeV. This is done by setting {\tt xptj = 150} in {\tt MadGraph5\_aMC@NLO}. We have checked that this setup doesn't lose the generality, but improves the event generating efficiency by a factor of $\sim6$. The background cross section after cuts is 2.88 pb taken into account the $K$-factor.

\begin{figure}
\centering
\subfigure[~Examples of distributions in the resolved region.]{
\label{fig:low_resolved}\includegraphics[scale=0.2]{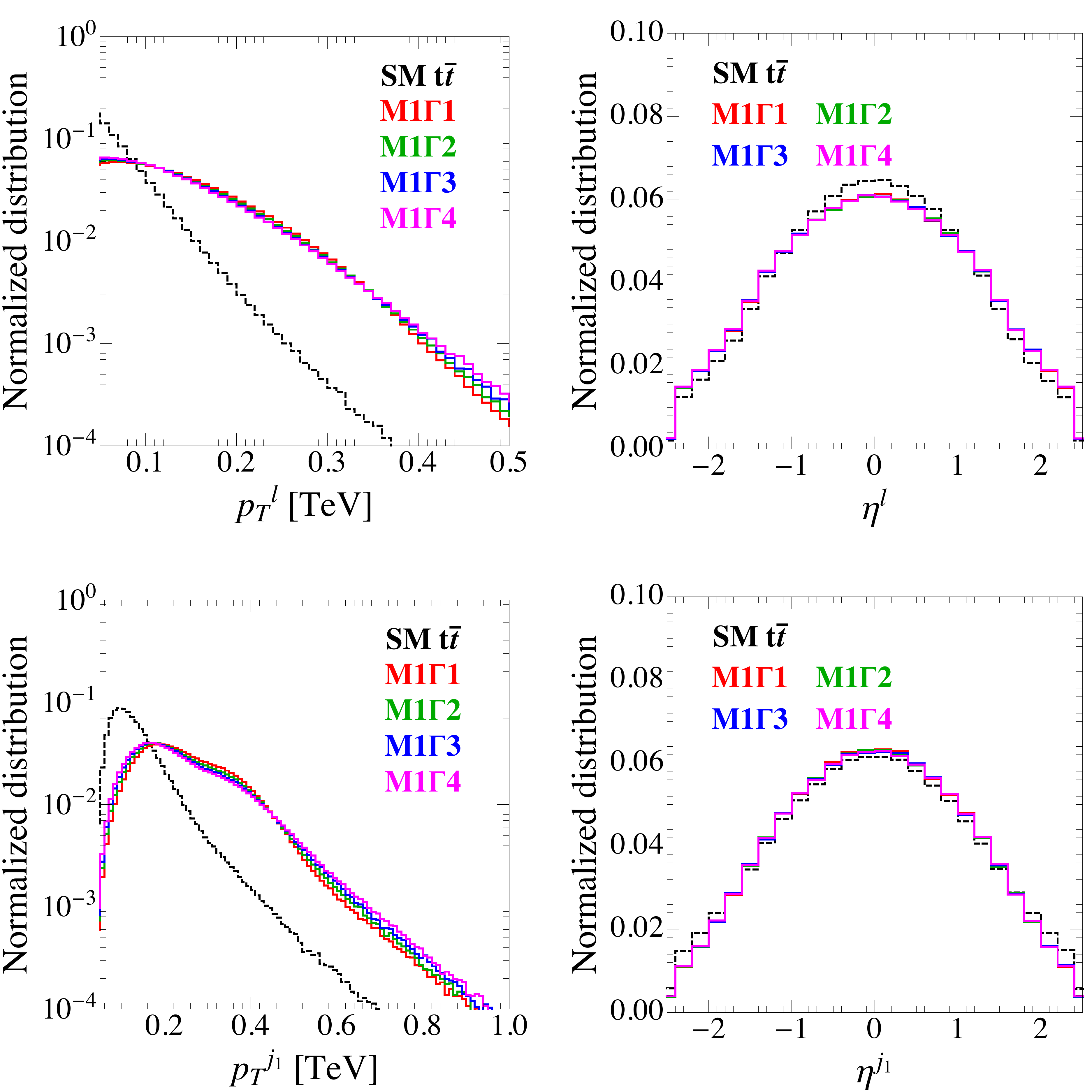}}
\subfigure[~Examples of distributions in the boosted region.]{
\label{fig:low_boosted}\includegraphics[scale=0.2]{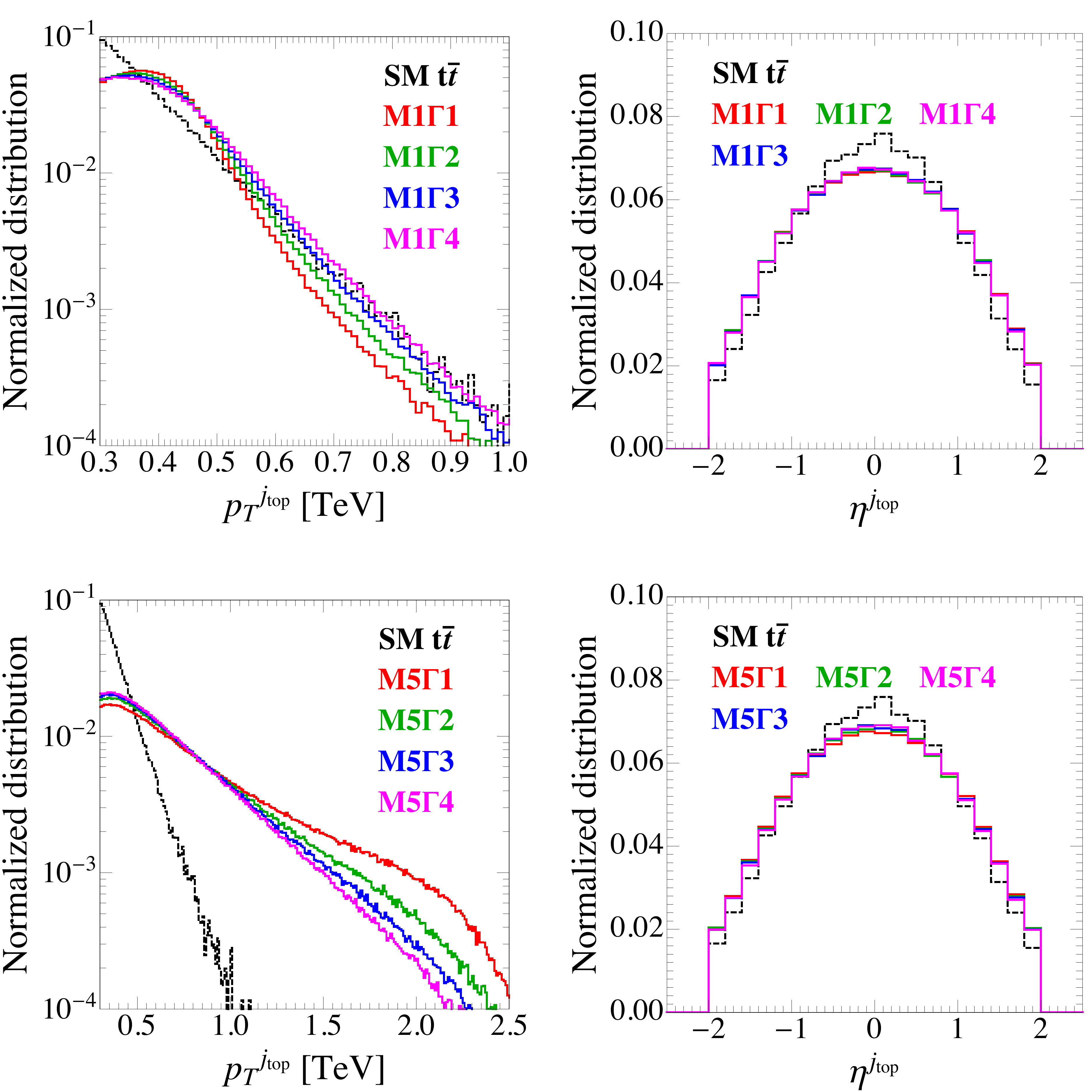}}
\caption{Distributions of some low-level observables used to train a DNN. Selection cuts in Table~\ref{tab:cut_flow_resolved} and \ref{tab:cut_flow_boosted} are applied.}
\label{fig:low_observables}
\end{figure}

The events after cuts are collected to make training and validation/test datasets. For the resolved region, we have $1\ell^\pm+\slashed{E}_T+4$ jets in total 6 reconstructed objects in the final state, and 26 low-level kinematic observables can be used as input features: $E^\ell$, $p_T^\ell$, $\eta^\ell$ and $\phi^\ell$ from the charged lepton; $\slashed{E}_T$, $\phi^{\slashed{E}_T}$ from the missing transverse momentum; $E^{j_i}$, $p_T^{j_i}$, $\eta^{j_i}$, $\phi^{j_i}$ and $b^{j_i}$ from the 4 leading jets, with $i=1$, 2, 3, 4. Here $b^j$ is the $b$-tagging observable, which is 1 for a $b$-tagged jet and 0 otherwise. Some examples of the low-level observables distributions are shown in Fig.~\ref{fig:low_resolved}. For each benchmark case (i.e. M1$\Gamma$1$\sim$M1$\Gamma$4), we build a training dataset and a validation/test dataset. Both of those two datasets have 1,000,000 events, which contain nearly equal signal and background events.

For the boosted region, $1\ell^\pm+\slashed{E}_T+1~\text{top-jet}+1~{\rm selected~jet}$ in total 4 objects are reconstructed, and we can extract 15 low-level observables as input features: the first 6 are from $\ell$ and $\slashed{E}_T$, same as the resolved region; the other 9 insist of $E^{j_{\rm sel}}$, $p_T^{j_{\rm sel}}$, $\eta^{j_{\rm sel}}$, $\phi^{j_{\rm sel}}$, $b^{j_{\rm sel}}$ from the selected jet, and $E^{j_{\rm top}}$, $p_T^{j_{\rm top}}$, $\eta^{j_{\rm top}}$, $\phi^{j_{\rm top}}$ from the top-jet. Some examples of the low-level observables distributions are illustrated in Fig.~\ref{fig:low_boosted}. For each benchmark case (i.e. M1$\Gamma$1$\sim$M1$\Gamma$4, and M5$\Gamma$1$\sim$M5$\Gamma$4), we randomly mix equal number of signal and background events to get 800,000 events for training and another 800,000 events for validation/test. 

\subsection{Training the DNN}\label{sec:training}

The DNN classifier is implemented using the {\tt Keras}~\cite{chollet2015keras} package (with {\tt Tensorflow}~\cite{abadi2016tensorflow} as the backend). The architecture of the DNN is as follows, 
\bea\label{DNN_structure}
[N_{\rm in},\underbrace{N_{\rm node},N_{\rm node},\cdots,N_{\rm node}}_{N_{\rm hidden}},2],
\eea
where $N_{\rm hidden}$ and $N_{\rm node}$ are the numbers of hidden layers and the number of neurons per hidden layer, respectively. The number of input features $N_{\rm in}=26$ (15) for the resolved (boosted) region. All the input features are rescaled to have average 0 and standard deviation 1 before training. We label the events with column matrices to match the two neurons in output layer:
\bea\label{label_df}
\text{signal}\to\begin{bmatrix}0\\ 1\end{bmatrix},~\text{background}\to\begin{bmatrix}1\\ 0\end{bmatrix}.
\eea
The Rectified Linear Unit ({\tt ReLU}) activation function is used for all the hidden layers, while the {\tt softmax} activation function is adopted for the output layer. The loss function is {\tt categorical\_crossentropy}, and the optimizer is {\tt Adam}. To get the best configuration of the DNN, we try various choices of the hyper-parameter combination as follows,
\begin{align}\label{eq:configurations}
&N_{\rm hidden}=4,~5;&N_{\rm node}=200,~300;\nn\\
&L_r=0.001,~0.003;&D_r=0.1,~0.2,~0.3;\nn\\
&B_s=10^3,~10^4;&
\end{align}
where $L_r$ is the initial learning rate, $D_r$ is the dropout rate, and $B_s$ is the batch size. For each benchmark case, there are in total 48 different DNN configurations, in which we select the best one based on the learning curves with the following criteria:

\begin{enumerate}
\item If the validation/test accuracy curve achieves its maximum when crossing with the training accuracy curve, and meanwhile the validation/test loss curve reaches its minimum and crosses with the training curve, we select that configuration and cut the training at that epoch. This early stop is to prevent over-fitting.
\item If more than one configurations have the behaviors mentioned above, then we select the one with the higher validation/test accuracy and lower validation/test loss; if still there remain more than one networks, we choose the one with learning curves having less fluctuation.
\end{enumerate}

\noindent The details of training and the chosen configurations are listed in Tables~\ref{tab:best_network_1} and \ref{tab:best_network_5} of the Appendix. For the $M_\rho=1$ TeV models, the DNN can reach a classification accuracy of $\geqslant80\%$ in the resolved region and of $\geqslant65\%$ in the boosted region. While for the $M_\rho=5$ TeV case, the accuracy is $\geqslant76\%$ in the boosted region.

\begin{figure}
\centering
\subfigure{
\includegraphics[scale=0.2]{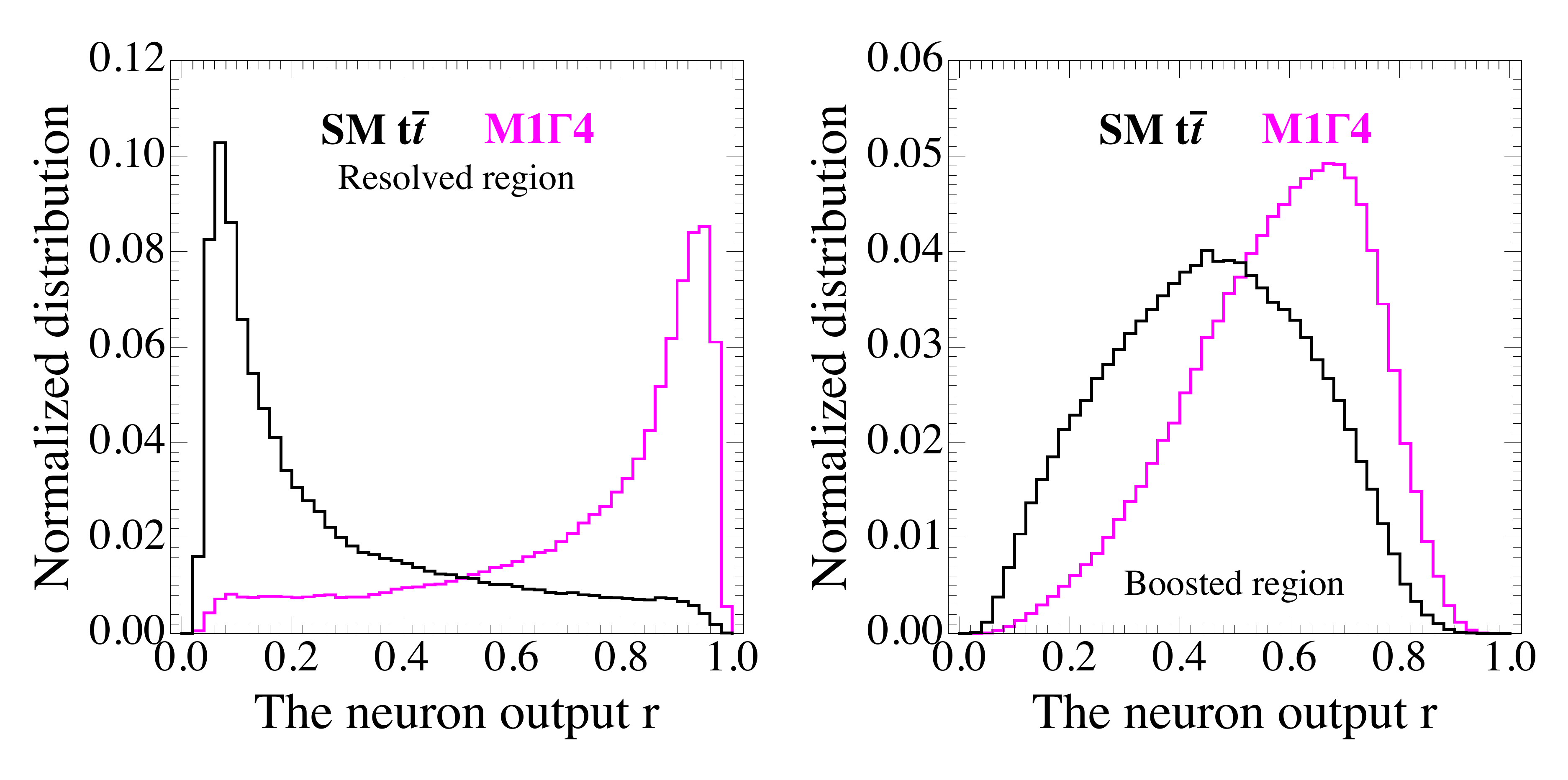}}
\subfigure{
\includegraphics[scale=0.2]{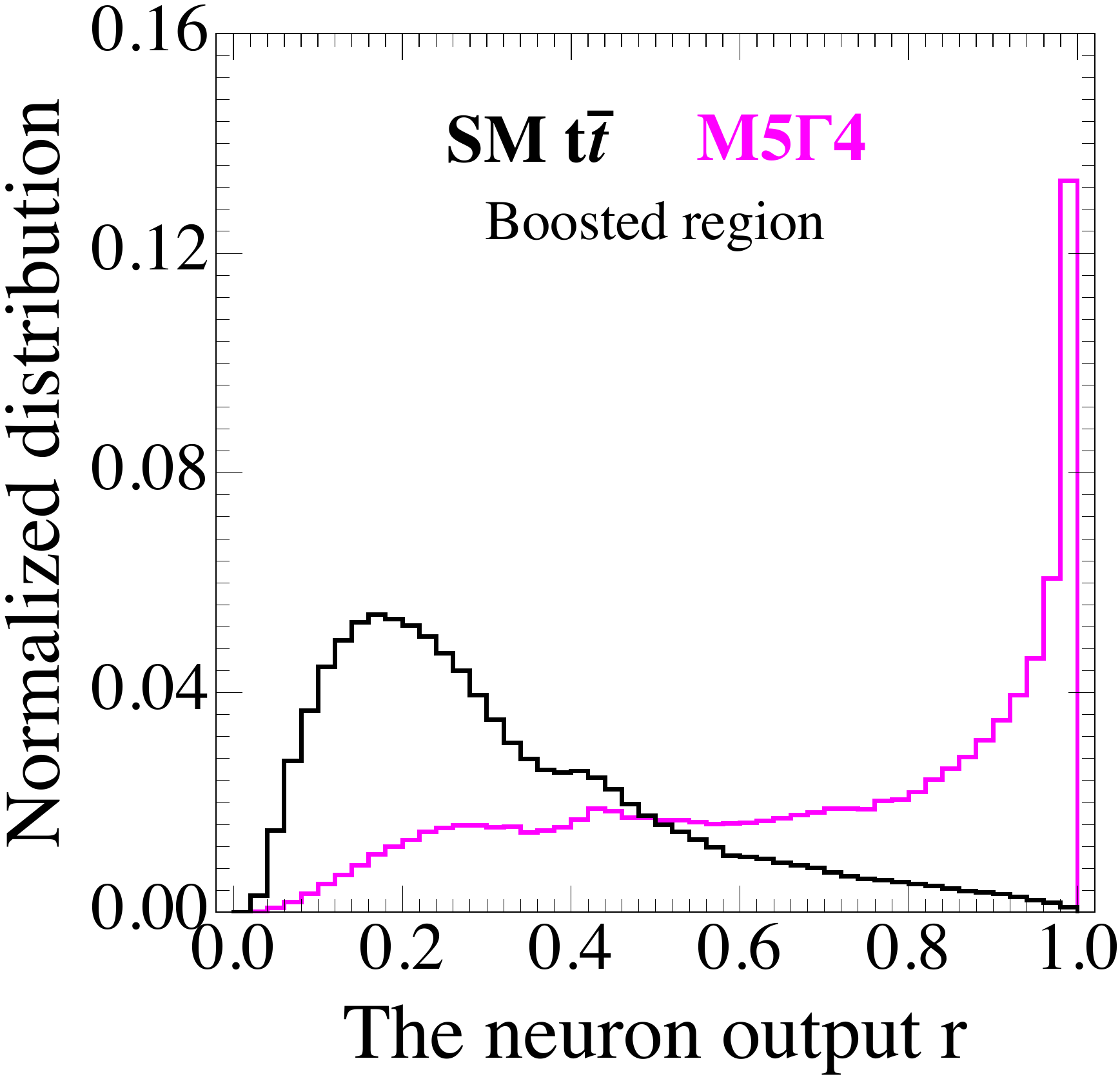}}
\caption{The final DNN output $r$ distributions that we use to obtain cross section upper limits. Benchmark cases with $\Gamma_\rho/M_\rho=40\%$ and $t\bar{t}$ backgrounds.  
   }
\label{fig:nn_response}
\end{figure}

The {\tt softmax} activation function for the output layer guarantees the output responses of the 0th neuron ($r_0$) and the 1st neuron ($r_1$) satisfy 
\be
0<r_0,\;r_1<1,\quad r_0+r_1\equiv 1.
\ee
Therefore, we can consider $r_1$ only, and denote it as $r$. Due to the label definition in \Eq{label_df}, If the DNN is well trained, the distribution of $r$ should have a peak around 1.0 (0.0) for the signal (background), for both the training data and the validation/test data. Figure~\ref{fig:nn_response} shows the distributions of the validation/test data for benchmark cases with $\Gamma_\rho/M_\rho=40\%$ as an illustration. The DNN for M1$\Gamma$4 shows worse performance in boosted region compare to the one in resolved region. This is because that two peaks in neuron output from signal and SM background are not separated well. In fact, this is a generic feature for all $M_\rho=1$ TeV benchmark cases. It is mainly due to the the boosted region cuts, which require a top-jet with $p_T^{j_{\rm top}}>300$ GeV. As a result, most of the SM $t\bar t$ background events are round this value. However, for a $M_\rho=1$ TeV resonance, its decay product $t/\bar t$ acquires a transverse momentum $\sim500$ GeV, quite similar to the cut threshold. Therefore, the signal and background look similar (see the $p_T^{j_{\rm top}}$ distribution in Fig.~\ref{fig:low_boosted}), and thus the  separation is not efficient. On the other hand, for a $M_\rho=5$ TeV resonance, $p_T^{j_{\rm top}}\sim2.5$ TeV, the DNN works very well, as plotted in the bottom of Fig.~\ref{fig:nn_response}.

\subsection{Setting bounds for the signal}

\begin{figure*}
\centering
\subfigure{
\includegraphics[scale=0.3]{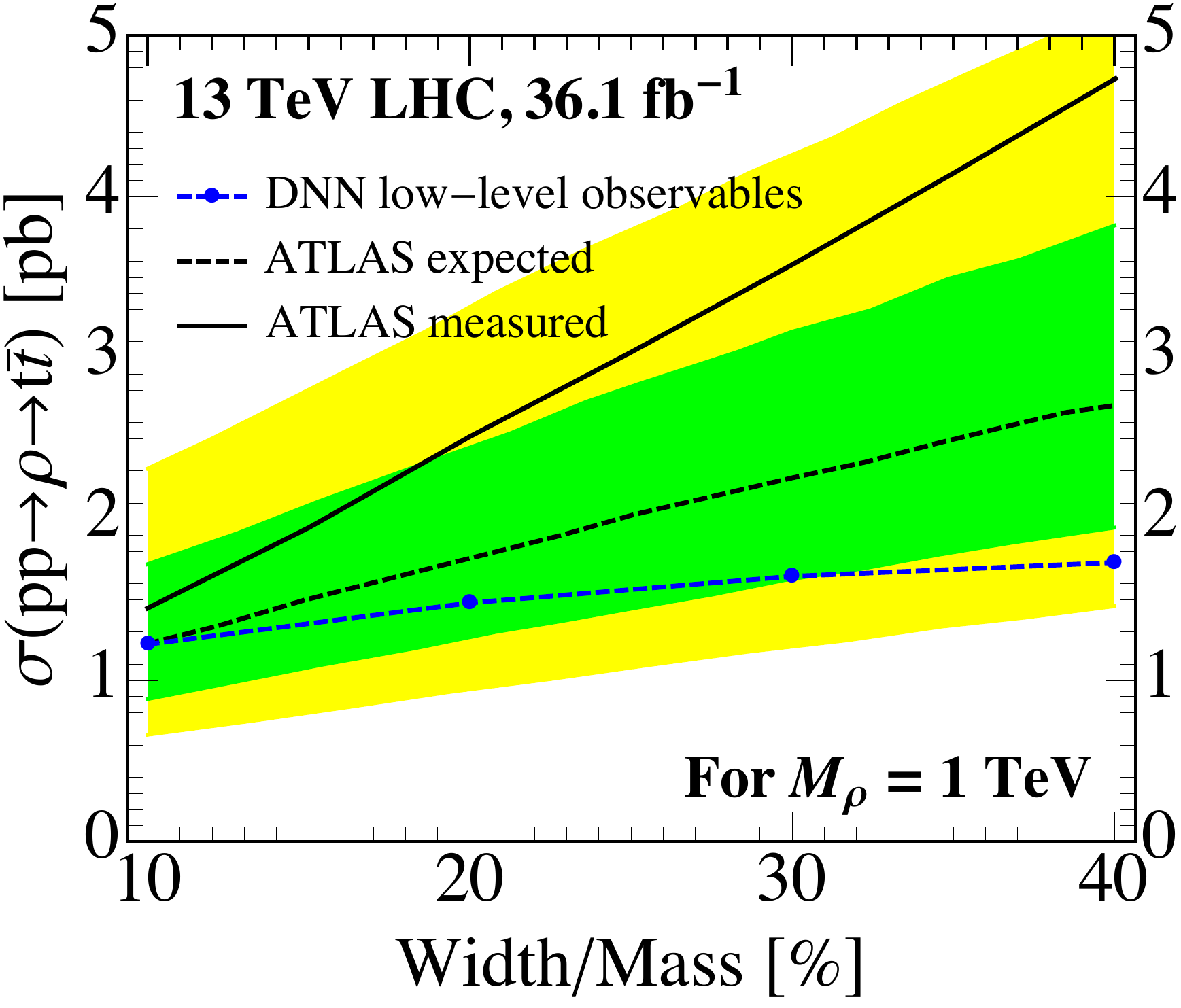}}\qquad\qquad\qquad
\subfigure{
\includegraphics[scale=0.29]{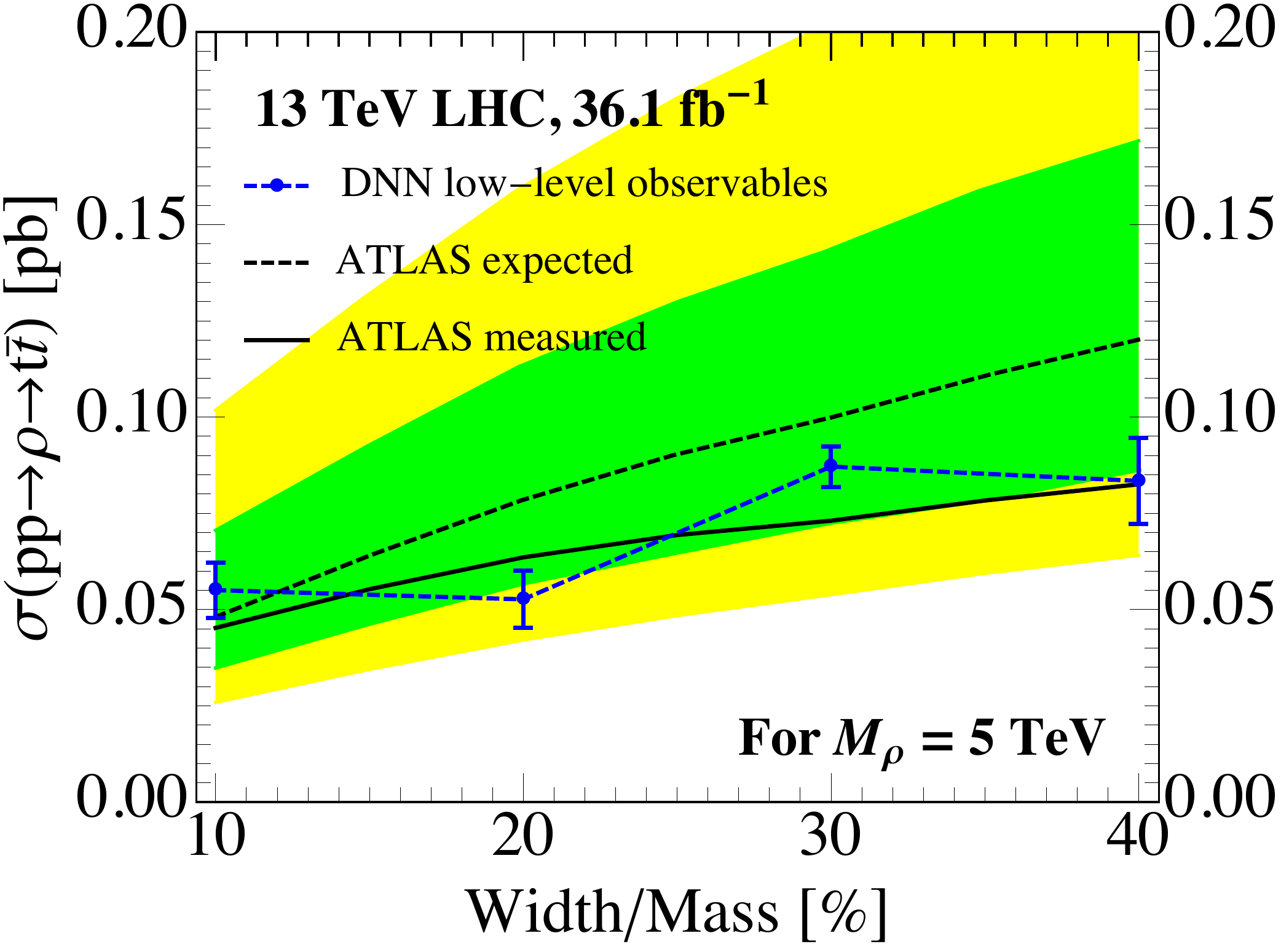}}
\caption{The DNN-improved cross section upper limits at 95\% C.L., obtained by fitting DNN output $r$ distributions in Fig.~\ref{fig:nn_response}. The latest ATLAS results~\cite{Aaboud:2018mjh} are also shown for comparison. The vertical error bars of the DNN results are training uncertainties, which are derived by running the same network for 15 times.}
\label{fig:cross_section}
\end{figure*}

We treat the neuron output $r$ as an observable, and fit its distribution shape to get the cross section upper limit of $pp\to\rho\to t\bar t$ for a given integrated luminosity. For the $M_\rho=1$ TeV benchmark cases, we use a binned $\chi^2$ fitting method by dividing the $0<r<1$ range into 50 bins. While for the $M_\rho=5$ TeV benchmarks, as the signal cross sections are expected to be tiny, to improve the efficiency we use the un-binned fitting method described in Refs.~\cite{Yang:2017iqh,CMS-NOTE-2011-005}. In each case, we consider the statistic uncertainty and assume a 12\% systematic uncertainty for the background. To include the effect of other subdominant backgrounds besides $t\bar t$ (i.e. $W+{\rm jets}$, multi-jet, etc), we further rescale the cross section by a factor of $1.23=1/0.81$ and $1.14=1/0.88$ for the resolved and boosted regions, respectively. Those factors come from the fact that $t\bar t$ contributes 81\% (88\%) of the total background for resolved (boosted) region~\cite{Aaboud:2018mjh}.
This simple rescaling could overestimate final contributions from subdominant backgrounds, and result in somewhat conservative estimations of cross section bounds.

The signal strength upper limits are derived for the unfolded parton-level cross section $\sigma(pp\to\rho\to t\bar t)$, which can be compared with the final results in experimental papers, e.g. Refs.~\cite{Aaboud:2018mjh,Sirunyan:2018ryr}. Our results are shown in Fig~\ref{fig:cross_section}, in which the expected and measured upper limits of Ref.~\cite{Aaboud:2018mjh} are also plotted as references, as they use the same final state and similar selection cuts. One can read that the DNN results are rather insensitive to the width of the $\rho$ resonance compare to the traditional approach, achieving better constraints in the large width region~\footnote{We also checked that the DNN results are better than those from more traditionally used BDT.}. For the $M_\rho=1$ TeV benchmark, the result is obtained by the combined fitting of both resolved and boosted regions. Individually, the resolved and boosted regions respectively yield cross sections $\sim3$ pb and $\sim1$ pb. Although networks in the resolved region have a higher accuracy ($\geqslant80\%$) than those in the boosted region ($\geqslant65\%$) in Table~\ref{tab:best_network_1}, they actually give a worse measurement of the cross section. This is because the boosted cuts can remove lots of background events and hence improve the fitting performance. That is also the reason why we only consider the boosted region for $M_\rho=5$ TeV: the production rate for such a high mass $\rho$ is so small that we have to use the boosted region to suppress the background. The DNN bounds for 5 TeV signal benchmark are comparable to the experimentally measured ones, but still better than the experimentally expected ones. As the training uses random number for the initialization of weights and biases, even for a given DNN configuration, the final results are slightly different for different running. To take into account this training uncertainty, we repeat 15 times of running the chosen DNN configuration for each benchmark case. For the $M_\rho=1$ TeV case, the relative fluctuation is small thus not shown; while for the $M_\rho=5$ TeV case, the standard deviations of the runs are shown as vertical error bars in Fig~\ref{fig:cross_section}.

\section{Figuring out what the machine had learned}\label{sec:learn}

In this section, we attempt to assess information learned by DNN using three methods, each of which will be discussed in each subsection. As a result, we can figure out not only which information has been learned, but also which information is most important.

\subsection{Testing high-level observables}

It is important to know whether a DNN had learned well-known useful but complicated features. In fact, it has been argued that some machine learning methods such as jet image~\cite{Cogan:2014oua} do not efficiently capture invariant mass features~\cite{deOliveira:2015xxd}.

Our approach is to train another set of DNNs using additional high-level observables, of which features we want to test. By comparing the performances of these new DNNs with the original DNNs trained with only low-level observables, we can test whether those particular high-level features (i.e. physically-motivated) have been effectively learned~\footnote{The definition of ``learned'' could be ambiguous, but we use subjective criteria discussed in text.} or not. This ``saturation approach'' has been widely used in particle physics research~\cite{Baldi:2014kfa,Datta:2017rhs}. 

To construct high-level observables, we first reconstruct the $t$ and $\bar t$. The longitudinal momentum of the neutrino is solved by requiring the leptonically decaying $W$ to be on-shell, i.e. $M_{\ell\nu}=M_W$. For the resolved region, the assignment of the 4 reconstructed jets are done by minimizing 
\begin{equation*}
\chi^2=\frac{(M_{jj}-M_W)^2}{\sigma_W^2}+\frac{(M_{jjj}-M_t)^2}{\sigma_t^2}
+\frac{(M_{j\ell\nu}-M_t)^2}{\sigma_t^2},
\end{equation*}
for various jet permutations, where $\sigma_W=0.1\times M_W$ and $\sigma_t=0.1\times M_t$. For the boosted region, a top quark is identified as the top-jet and the other is reconstructed from the combination of $\ell^\pm\nu j_{\rm sel}$. Once the $t$ and $\bar t$ are reconstructed, we are able to define the following 7 high-level observables for the signal $pp\to\rho\to t\bar t$:

\begin{enumerate}
\item The invariant mass $M_{t\bar t}$ of the $t\bar t$ system.
\item The polar angle and azimuthal angle in the Collins-Soper frame~\cite{Collins:1977iv}. We label the leptonic and hadronic decaying tops with subscripts ``tl'' and ``th'', respectively. Hence we have $\cos\theta^{\rm CS}_{\rm tl}$, $\cos\theta^{\rm CS}_{\rm th}$, $\phi^{\rm CS}_{\rm tl}$ and $\phi^{\rm CS}_{\rm th}$ in total 4 observables.
\item The polar angles in the Mustraal frame~\cite{Richter-Was:2016mal}, $\cos\theta_1^{\rm Mus.}$ and $\cos\theta_2^{\rm Mus.}$.
\end{enumerate}

\noindent 
The first observable reveals the resonance feature, while the latter 6 observables reflect the spin-1 nature of the $\rho$ resonance. For the boosted region, to take into account the features of the top jet, we introduce 3 additional high-level observables, i.e.
\begin{enumerate}
\item The invariant mass $M_{j_{\rm top}}$ of the top jet.
\item The $N$-subjettiness observables $\tau_{21}$ and $\tau_{32}$ of the top jet~\cite{Kaplan:2008ie,Thaler:2010tr,Thaler:2011gf,Plehn:2011tg,Kasieczka:2015jma}.
\end{enumerate}
Those observables are shown to be important in identifying the color structure of the hard process~\cite{Kasieczka:2015jma,Joshi:2012pu,Soper:2014rya}. In our scenario, the signal results from a color-singlet resonance, while the background comes from QCD process, and the jet mass and $N$-subjettiness can help to reveal this difference~\cite{Joshi:2012pu}.
Moreover, such jet substructures can be more independent on resonance characteristics and kinematics.

\begin{figure}
\centering
\subfigure[~Examples of distributions in the resolved region.]{
\includegraphics[scale=0.22]{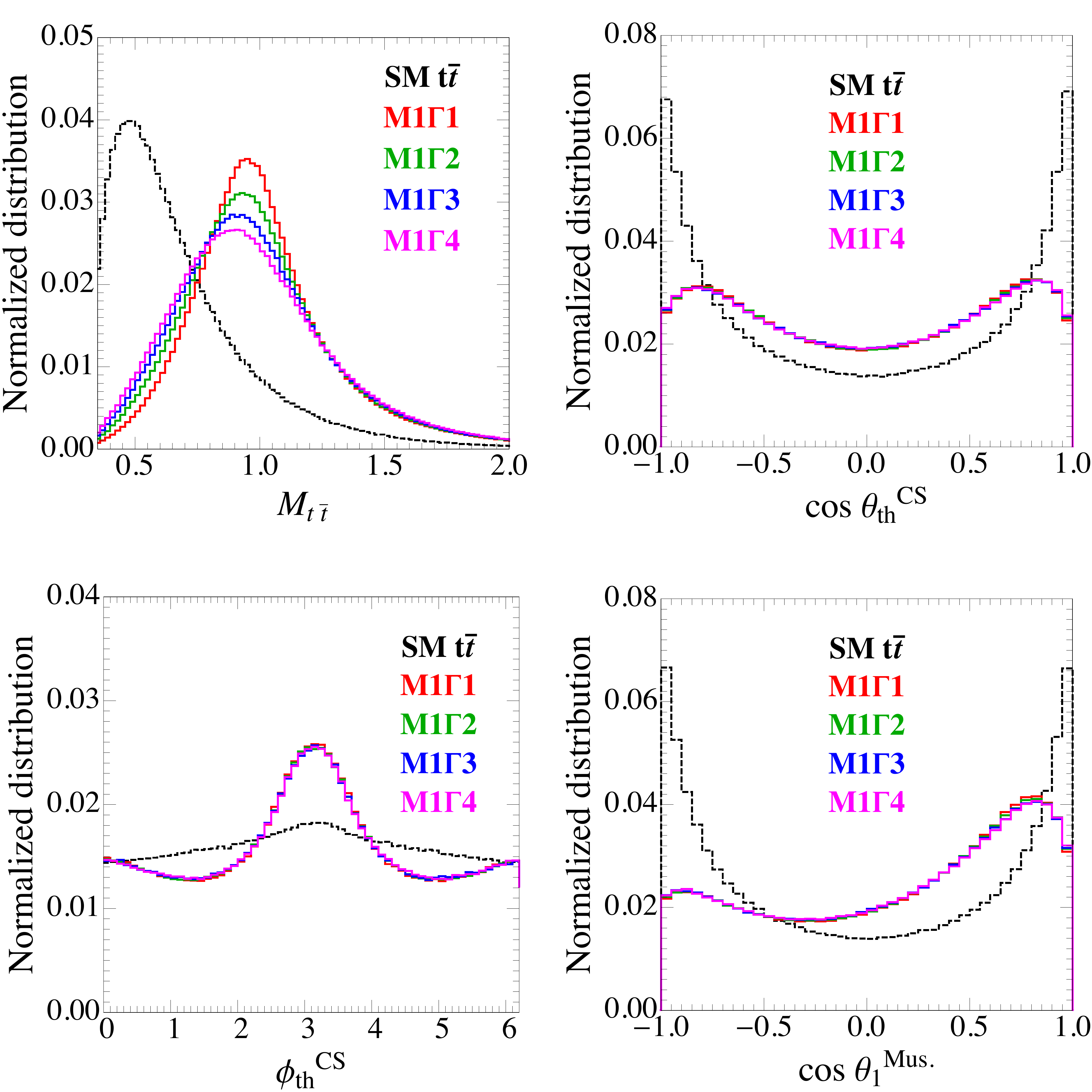}}
\subfigure[~Examples of distributions in the boosted region.]{
\includegraphics[scale=0.22]{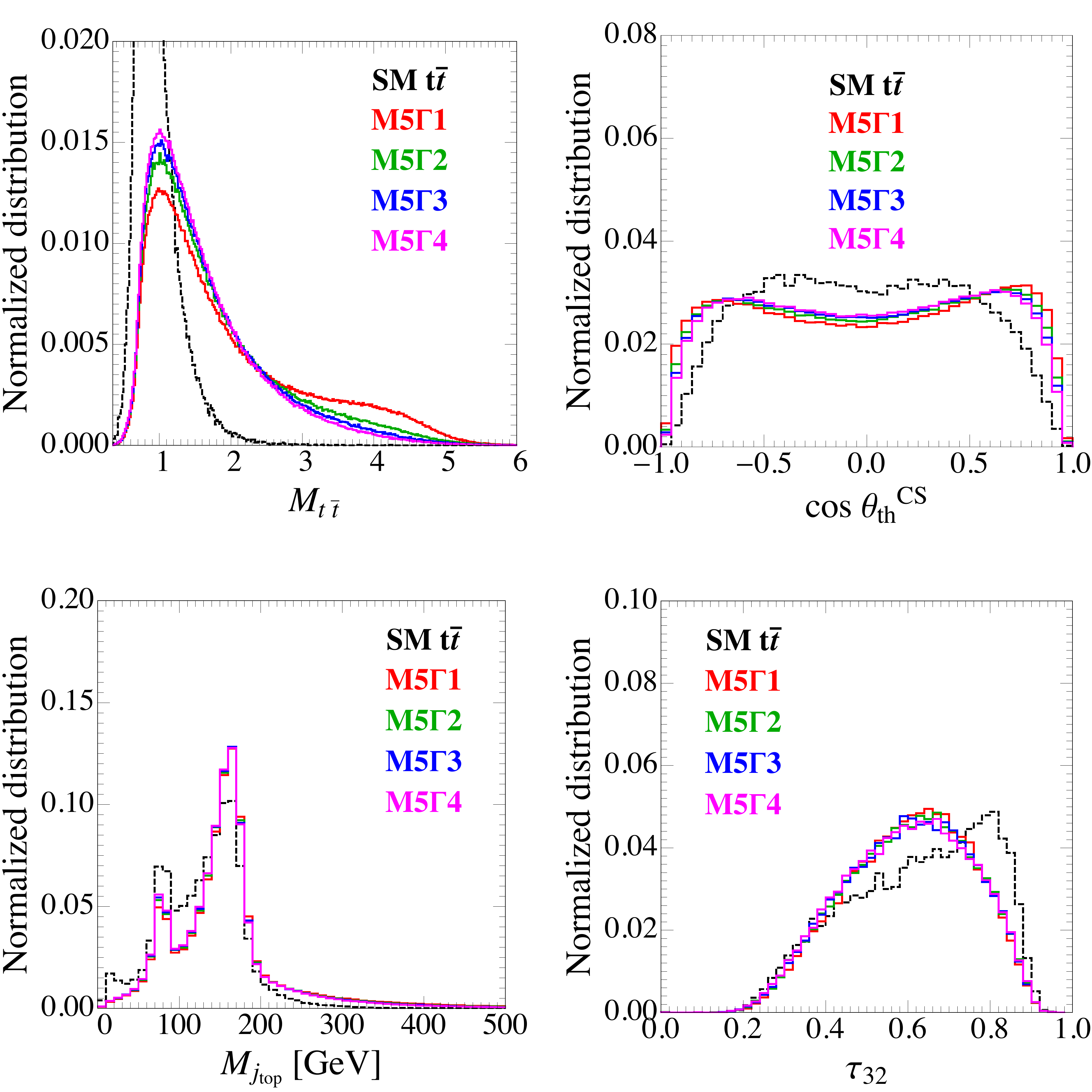}}
\caption{
Distributions of high-level observables in the (a) resolved and (b) boosted regions.} We use these to train a new set of DNNs to test whether such high-level features were learned.
\label{fig:high_observables}
\end{figure}

Some distributions of these high-level observables are shown in Fig.~\ref{fig:high_observables}. Note that the spin correlations as well as the jet substructure observables are rather insensitive to the width of $\rho$, as expected.
For the 5 TeV resonance, the mass peak of $M_{t\bar t}\sim5$ TeV almost disappears for $\Gamma_\rho/M_\rho\geqslant10\%$; instead, there is a peak $\sim1$ TeV, due to the parton-distribution support of off-shell effects and hard $p_T$ cuts. Most identified top-jets in both signal and background originate correctly from the top quark, thus the differences shown in the distributions of $M_{j_{\rm top}}$ and $\tau_{32}$ come from the color structure of the hard process. For example, the background's $M_{j_{\rm top}}$ distribution is slightly broader and the $\tau_{32}$ is slightly bigger than the signals. This is because the top-jets from QCD $t\bar t$ are color connected with the initial state, consequently having more radiations.
Using these ``all observables'' (i.e. sum of low- and high-level observables) as inputs, we train a new set of DNNs; best network configurations are again surveyed and detailed in Tables~\ref{tab:best_network_1} and \ref{tab:best_network_5} of the Appendix. 

\begin{figure}
\centering
\subfigure{
\includegraphics[scale=0.115]{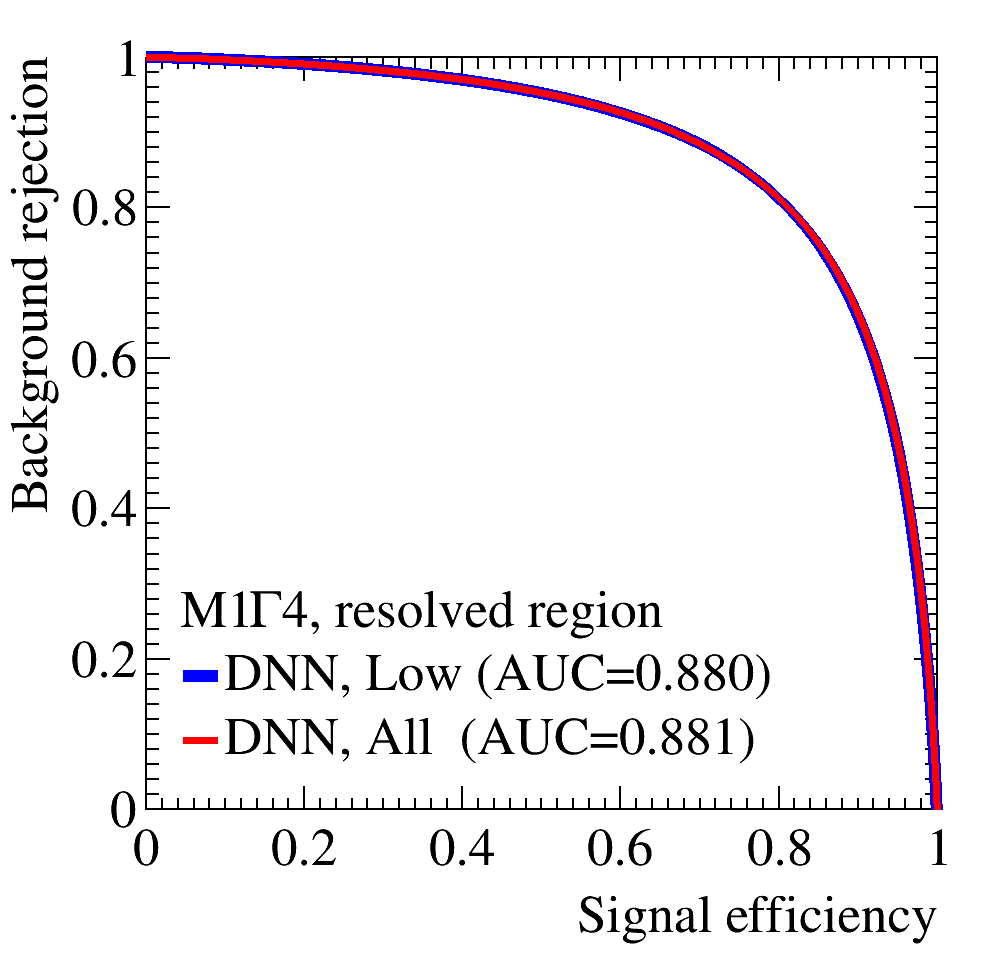}}\\
\subfigure{
\includegraphics[scale=0.115]{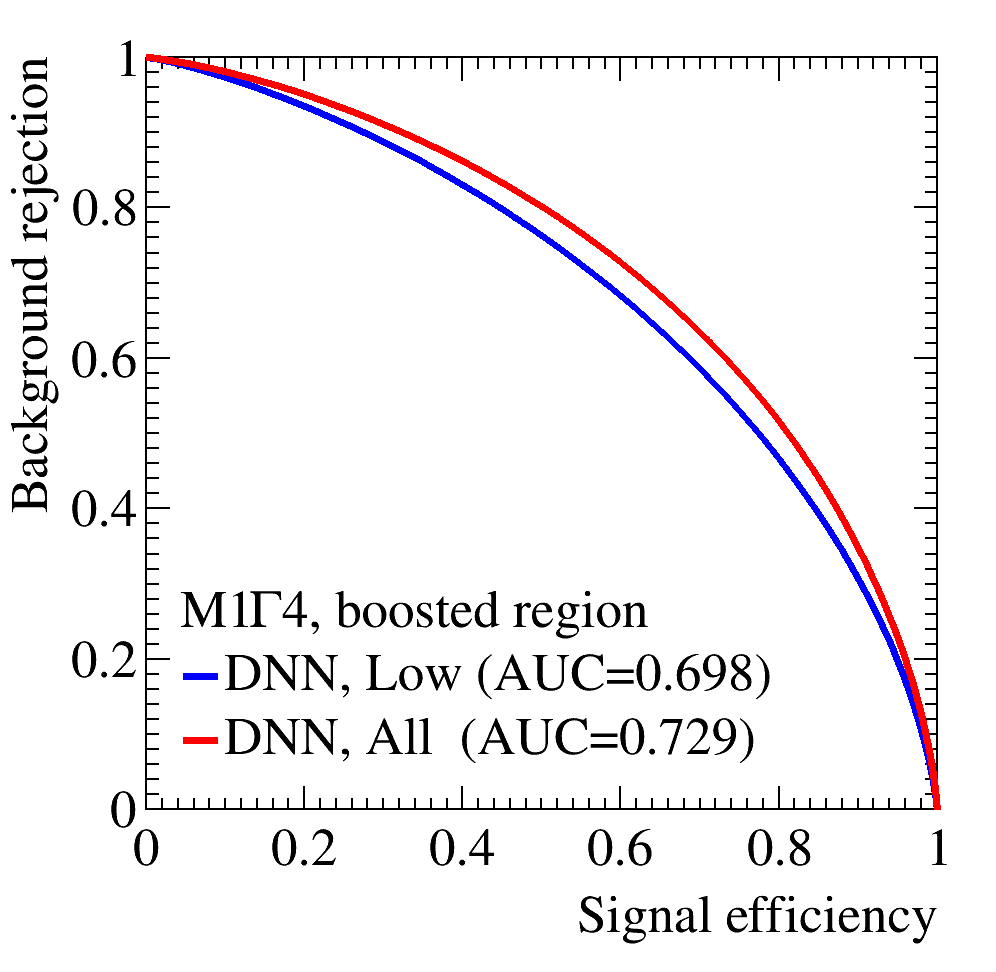}}\quad
\subfigure{
\includegraphics[scale=0.115]{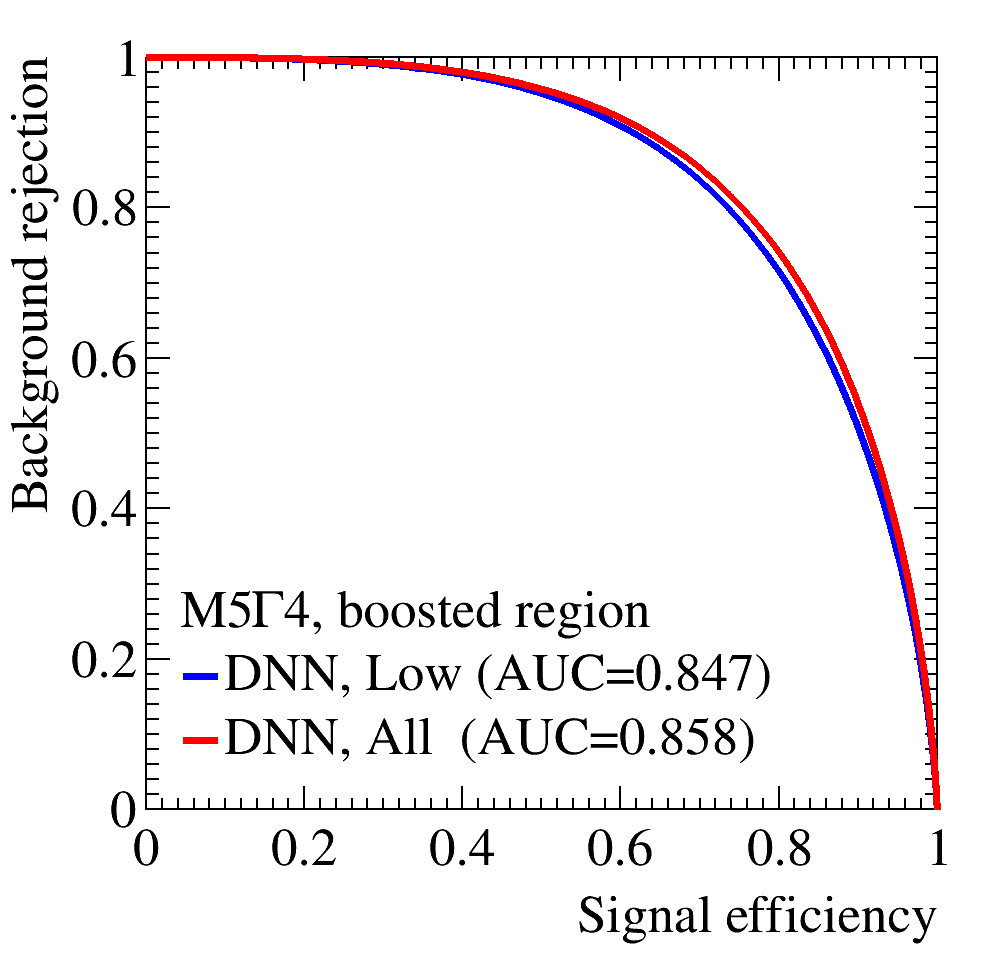}}
\caption{ROC curves, comparing the performance with (``all'') and without (``low'') high-level observables used to train DNNs. The AUC of each curve is also shown inside parenthesis.}
\label{fig:ROC}
\end{figure}

We compare the performances of original and new DNNs using receiver operating characteristic (ROC) curves. The area under curve (AUC) is used as a metric of the performance. Some of the comparisons are shown in Fig.~\ref{fig:ROC}. First, in the resolved region as shown in the top panel, we found that there is only little change on ROC curves by adding high-level observables. Not only AUC, but also background efficiencies show small change. This means that the inclusion of high-level observables does not yield the improvement of accuracy; the original DNN had learned those high-level features successfully from low-level inputs.

In the boosted region, while the $M_{t\bar t}$, $M_{j_{\rm top}}$ and spin correlations can be derived from the four momenta of reconstructed objects, the $N$-subjettiness cannot be inferred from the low-level inputs. Therefore, adding high-level features can bring improvements. As shown with ROC in the bottom two panels of Fig.~\ref{fig:ROC}, the improvement is sizable for M1$\Gamma$4, while, however, relatively small for M5$\Gamma$4. This may be because the event topology of M5 boosted cases becomes so simple that many features are more correlated.

\subsection{Ranking input observables by importance}

\begin{figure*}
\centering
\subfigure{
\includegraphics[scale=0.45]{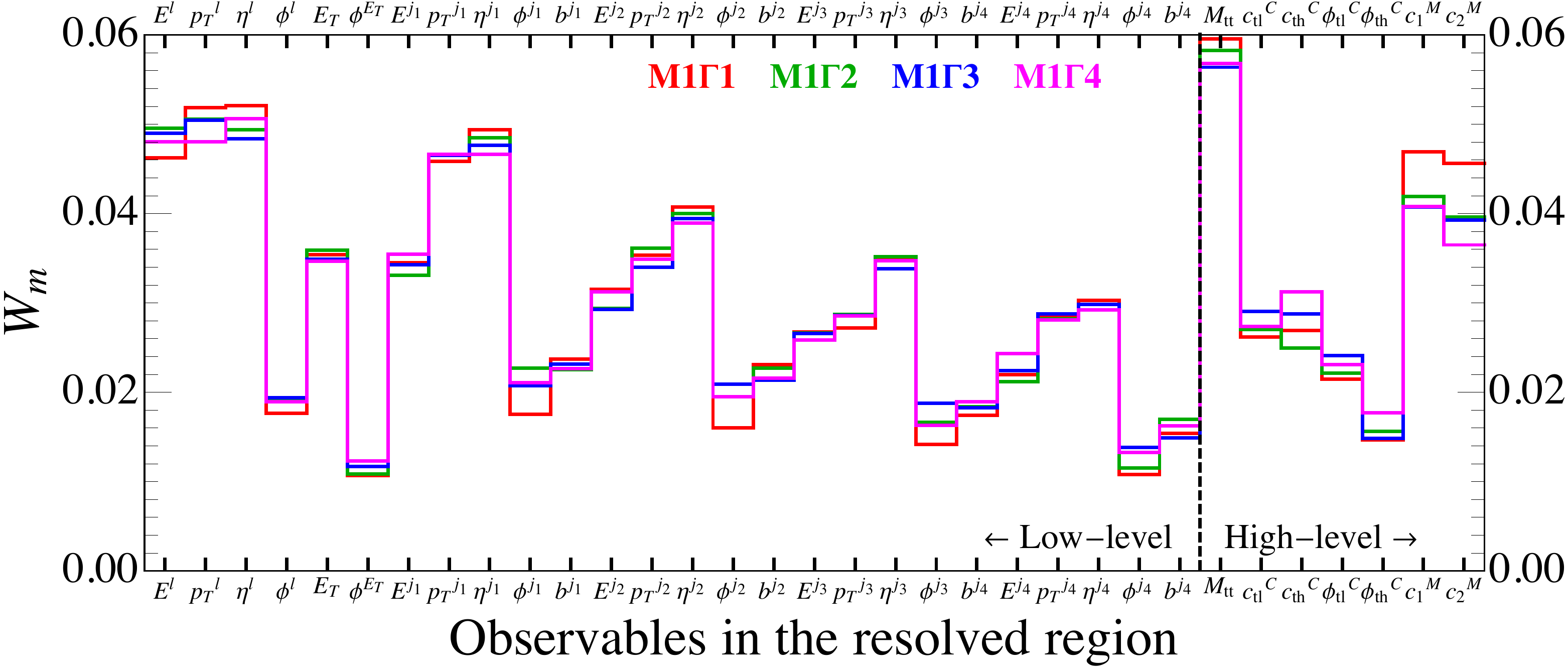}}\\
\subfigure{
\includegraphics[scale=0.45]{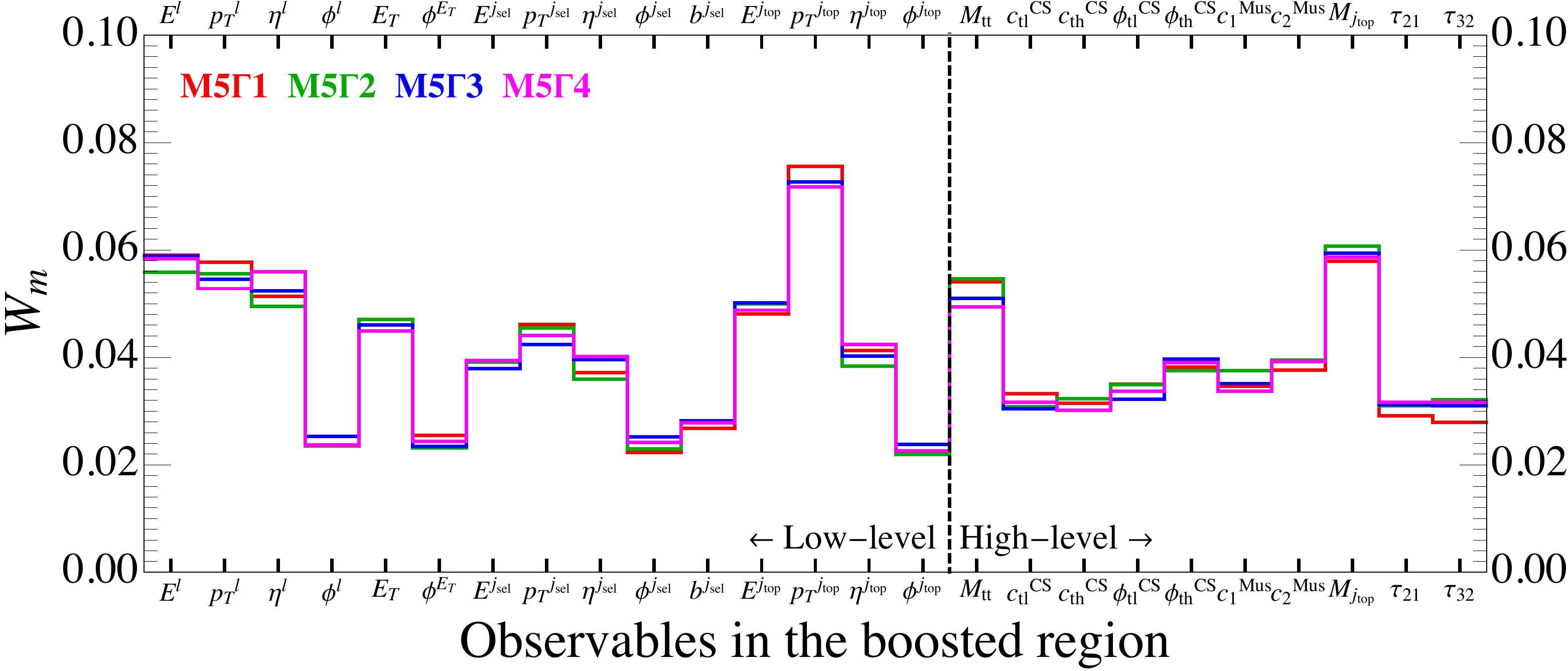}}\\
\caption{The weight of each observable in the first layer of DNN, $W_m$ defined in \Eq{eq:wm}. $M_\rho=1$ TeV in the resolved region (upper panel) and $M_\rho=5$ TeV in the boosted region (lower panel). High-level observables are also used in training in order to get those weights. For notations, $E_T$ and $\phi^{E_T}$ in the figures stand for $\slashed{E}_T$ and $\phi^{\slashed{E}_T}$, respectively. $c_{\rm tl,th}^{\rm C}$ and $c_{\rm tl,th}^{\rm CS}$ are short for $\cos\theta_{\rm tl,th}^{\rm CS}$, and $c_{1,2}^{\rm M}$ and $c_{1,2}^{\rm Mus}$ for $\cos\theta_{1,2}^{\rm Mus.}$.}
\label{fig:weights}
\end{figure*}

Which information has been used most usefully by DNN in distinguishing a broad resonance against continuum background? To answer this, we attempt to identify which connections between which neurons and layers are weighted most importantly. Following
Ref.~\cite{nielsen2015neural}, we define the learning speed of the $j$-th hidden layer as
\bea
v^{(j)}=\left|\frac{\partial\mL_{\rm loss}(w,b)}{\partial \vec{b}^{(j)}}\right|,
\eea
where $\vec{b}^{(j)}$ is the bias vector of the $j$-th hidden layer, while $\mL_{\rm loss}$ is the loss function. As the target of machine learning is to find the global minimum of $\mL_{\rm loss}$, the $v^{(j)}$ approximately reflects the training sensitivity of a specific layer. When training the DNN, the larger $v^{(j)}$ a layer acquires, the more important it is. We found that for all individual benchmark cases M$i\Gamma j$, the first hidden layer has the highest learning speed several times larger than that of other layers. For example, for M1$\Gamma$4 case in the resolved region,
  the learning speed is $v^{(1)}=0.457$, $v^{(2)}=0.086$, $v^{(3)}=0.033$, $v^{(4)}=0.016$ and $v^{(5)}=0.008$. This means that good features are typically learned most efficiently in the first hidden layer.

For our DNN architecture described in \Eq{DNN_structure}, the weights of the first hidden layer form a $N_{\rm in}\times N_{\rm node}$ matrix, whose element is denoted as $w^{(1)}_{mn}$ with $m=1,\cdots,N_{\rm in}$ and $n=1,\cdots,N_{\rm node}$. As all the input features are rescaled to have average 0 and standard deviation 1, the magnitude of the weight $w^{(1)}_{mn}$ reflects the correlation strength between the $m$-th input and the $n$-th neuron in the first hidden layer. Motivated by this, we further define
\bea
W_m = \mN\sqrt{\sum_{n=1}^{N_{\rm node}}\left(w^{(1)}_{mn}\right)^2},
\label{eq:wm} \eea
as a measure of the importance of the $m$-th input feature. The normalization $\mN$ is such that
\bea
\sum_{m=1}^{N_{\rm in}} W_m = 1.
\eea

\medskip
Figure~\ref{fig:weights} shows the $W_m$'s of each input observable from the DNN trained using both low- and high-level observables. Above all, the $M_{t\bar{t}}$ -- that we expected to be less useful for a broad resonance -- is still one of the most important observables even when the resonance is broad. This is particularly true for a low-mass broad resonance in the resolved region (upper panel). In the case of a heavy-resonance in the boosted region (lower panel), its importance is relatively reduced, partly because some invariant-mass information has been used in the selection of the boosted region. In such cases, the top-jet mass and transverse momentum which are somewhat correlated with $M_{t\bar t}$ and width can significantly complement the search, as shown in the bottom panel.
In addition, the invariant mass of the top-jet is another important input feature because it reflects the color flow difference between signals and background. On the other hand, $N$-subjettinesses again turn out to be relatively less useful.

Remarkably, there are much other useful information, particularly from angular distributions $\eta^{\ell,j}$ and $\cos\theta_{1,2}^{\rm Mus.}$. From Figs.~\ref{fig:low_observables} and \ref{fig:high_observables}, we can see that these observables are relatively  uncorrelated with the resonance width. We have indeed checked that the cross entropies~\cite{Roxlo:2018adx} between these observables and $M_{t\bar t}$, which can quantify their correlations, are not so high. As we will see in the next subsection, these information are useful even in the off-shell region away from the resonance, hence less correlated with the width. Thus, these features are useful in search of broad resonances. This may also imply that narrow-resonance searches can be improved by adding off-resonance information; this is partly because a large fraction of signals is still from low-energy off-resonance region where parton-luminosity support is much larger (although buried under larger backgrounds). We leave this for a future study.

\subsection{Planing away $M_{t \bar t}$}\label{sec:planing}

We have observed that $M_{t\bar t}$ is still important, but there are indeed uncorrelated useful information. How much is discovery capability attributed to those uncorrelated (whether known or unknown) information?  Using the data planing method~\cite{deOliveira:2015xxd,Chang:2017kvc}, we plane away the feature in the invariant mass spectrum. We attach a weight to each event so that the weighted distribution of $M_{t\bar t}$ becomes flat for both signals and backgrounds; the details of chosen network configurations and more results are described in Table \ref{tab:planing} of the Appendix. A new set of DNNs trained with such planed data must learn information uncorrelated with $M_{t\bar t}$, and the difference between the performance with/without $M_{t\bar t}$ offers a quantitative answer to the question ``how much information it is beyond the invariant mass''.

In practice, to avoid large fluctuations, we use only $M_{t\bar t}\in[0.5,3]$ TeV region with 20 GeV bin size for all signal cases. This means that for 5 TeV signals, we consider only off-resonance events; note that the majority of signal is from the low-energy region supported by larger parton luminosities. 

After $M_{t\bar t}$ planed away, the classification accuracies reduce from $\geqslant80\%$ to $\geqslant73\%$ for $M_\rho=1$ TeV in the resolved region and from $\geqslant65\%$ to $\geqslant62\%$ in the boosted region. For $M_\rho=5$ TeV cases, accuracies reduce from $\geqslant76\%$ to $\geqslant63\%$ in the boosted region. As accuracies are still significantly higher than random guess (i.e. 50\%), we conclude that DNNs still have some capabilities to distinguish signals from background, even though they are blind to $M_{t\bar t}$ and most events are from off-resonance region (for 5 TeV cases). Clearly, on top of $M_{t\bar t}$ and width, the original DNNs had learned extra information (such as aforementioned angular correlations).

Indeed, we have checked that the weights $W_m$ for various anglular and angular-correlation observables, after planing the $M_{t\bar t}$, are relatively high. From Fig.~\ref{fig:low_observables} and \ref{fig:high_observables}, one can also see that they are largely independent on the width. The helicity conservation (hence, angular correlations) can hold somewhat independently of the invariant mass, as the range of the invariant mass considered is always much larger than the top mass. Thus, we conclude that much of the angular information can be from off-resonance region, and such off-resonance information (although buried under larger backgrounds) can enhance discovery power. As a result, as shown in Fig.~\ref{fig:cross_section}, final performance is not only improved but became rather insensitive to the resonance width.

A final remark is that there could still be {\it unknown} (to us) useful information that are not identified in our analysis.

\section{Conclusion}\label{sec:conclusion}

We have found that, in an attempt to develop methods to discover broad $t\bar t$ resonances, $M_{t\bar t}$ is still one of the most important observables, but additional information from both on- and off-resonance regions can significantly enhance discovery capability. As a result, the cross section upper limits can be improved by $\sim60\%$ for $\Gamma_\rho / M_\rho \sim 40\%$, and the improved LHC sensitivities do not strongly depend on the width of a resonance. As resonances in new physics beyond the SM are easily broad, our learnings and techniques can be used to efficiently search for them.

The most useful observables turn out to be $M_{t\bar t}$ (even for broad resonances), $p_T^{j_{\rm top}}$, $M_{j_{\rm top}}$, angular distributions and color correlations. The usefulness of $M_{t\bar t}$ even for broad-resonance searches is not necessarily obvious, {\it a priori}. But correlated observables such as $p_T^{j_{\rm top}}$ are found to further complement. Angular information (some of whose contributions come from off-resonance region) and $M_{j_{\rm top}}$ (which can measure color flow structures irrespective of resonance characteristics) are relatively uncorrelated with the width and $M_{t\bar t}$, making improved LHC sensitivities less dependent on the width.  Lastly, as we trained using only low-level inputs, our results also show that high-level observables such as $M_{t\bar t}$ are effectively well learned by DNN.

We have assessed these machine-learned information in three ways: by explicitly testing those high-level observables, by ranking input (low and/or high) observables using weights of the network, and by planing away features correlated with $M_{t\bar t}$. Notably, after all, there can still be unknown useful information that are not easily identified in our analysis. Thus, being able to communicate more efficiently with networks will enable better explorations of the nature, beyond what we know.

\begin{acknowledgements}
We would like to thank Shawn Jia, Jinmian Li, Hui Luo, Tao Xu, Daneng Yang and Zhao-Huan Yu for discussions and and the anonymous referee for useful suggestions. SJ and KPX are supported by Grant Korea NRF 2015R1A4A1042542, NRF 2017R1D1A1B03030820, SJ also by POSCO Science Fellowship, and DL by NRF 0426-20170003, NRF 0409-20190120.
\end{acknowledgements}

\appendix
\section{The chosen DNN configurations and their performances}\label{app:configuration}

\begin{table*}
\scriptsize\renewcommand\arraystretch{1.5}\centering
\begin{tabular}{|c|c|c|c|c|c|c|c|c|}\hline
Benchmark case & Kinematic region & Observables & $N_{\rm hidden}$, $N_{\rm node}$, $L_r$, $D_r$, $B_s$, $N_{\rm epoch}$ & Classification accuracy \\ \hline
\multirow{4}{*}{M1$\Gamma$1} & \multirow{2}{*}{resolved} & low-level & 5, 200, 0.001, 0.2, $10^3$, 150 & 85.2\% \\ \cline{3-5}
 &  & all & 5, 200, 0.001, 0.2, $10^3$, 100 & 85.1\% \\ \cline{2-5}
 & \multirow{2}{*}{boosted} & low-level & 5, 200, 0.001, 0.2, $10^4$, 55 & 67.9\%\\ \cline{3-5}
 & & all & 4, 300, 0.001, 0.3, $10^3$, 40 & 70.1\% \\ \hline
\multirow{4}{*}{M1$\Gamma$2} & \multirow{2}{*}{resolved} & low-level & 4, 300, 0.003, 0.2, $10^3$, 35 & 83.2\% \\ \cline{3-5}
 &  & all & 5, 200, 0.001, 0.2, $10^3$, 60 & 83.2\% \\ \cline{2-5}
 & \multirow{2}{*}{boosted} & low-level & 5, 200, 0.001, 0.2, $10^4$, 45 & 65.8\% \\ \cline{3-5}
 & & all & 4, 300, 0.003, 0.3, $10^4$, 40 & 68.2\% \\ \hline
\multirow{4}{*}{M1$\Gamma$3} & \multirow{2}{*}{resolved} & low-level & 4, 300, 0.001, 0.2, $10^3$, 30 & 81.6\% \\ \cline{3-5}
 &  & all & 5, 200, 0.001, 0.2, $10^3$, 40 & 81.6\%\\ \cline{2-5}
 & \multirow{2}{*}{boosted} & low-level & 4, 300, 0.003, 0.2, $10^4$, 30 & 65.1\% \\ \cline{3-5}
 & & all & 4, 300, 0.003, 0.3, $10^4$, 40 & 67.0\% \\ \hline
\multirow{4}{*}{M1$\Gamma$4} & \multirow{2}{*}{resolved} & low-level & 5, 200, 0.001, 0.2, $10^3$, 80 & 80.8\% \\ \cline{3-5}
 &  & all & 5, 200, 0.001, 0.2, $10^3$, 40 & 80.6\%\\ \cline{2-5}
 & \multirow{2}{*}{boosted} & low-level & 4, 300, 0.001, 0.2, $10^4$, 20 & 64.3\% \\ \cline{3-5}
 & & all & 4, 300, 0.001, 0.3, $10^3$, 40 & 66.7\% \\ \hline
\end{tabular}
\caption{The selected networks for $M_\rho=1$ TeV. $N_{\rm epoch}$ is the epoch number when we cut the training.}
\label{tab:best_network_1}
\end{table*}

\begin{table*}
\scriptsize\renewcommand\arraystretch{1.5}\centering
\begin{tabular}{|c|c|c|c|c|c|c|c|c|}\hline
Benchmark case & Kinematic region & Observables & $N_{\rm hidden}$, $N_{\rm node}$, $L_r$, $D_r$, $B_s$, $N_{\rm epoch}$ & Classification accuracy \\ \hline
\multirow{2}{*}{M5$\Gamma$1} & \multirow{2}{*}{boosted} & low-level & 4, 300, 0.001, 0.2, $10^3$, 20 & 79.5\% \\ \cline{3-5}
 & & all & 5, 200, 0.001, 0.1, $10^4$, 30 & 80.5\% \\ \hline
 \multirow{2}{*}{M5$\Gamma$2} & \multirow{2}{*}{boosted} & low-level & 4, 300, 0.003, 0.2, $10^3$, 40 & 78.2\%  \\ \cline{3-5}
 & & all & 5, 200, 0.001, 0.1, $10^4$, 45 & 79.1\% \\ \hline
 \multirow{2}{*}{M5$\Gamma$3} & \multirow{2}{*}{boosted} & low-level & 4, 300, 0.003, 0.2, $10^3$, 30 & 77.4\% \\ \cline{3-5}
 & & all & 4, 300, 0.003, 0.3, $10^4$, 40 & 78.4\% \\ \hline
 \multirow{2}{*}{M5$\Gamma$4} & \multirow{2}{*}{boosted} & low-level & 4, 300, 0.003, 0.2, $10^3$, 45 & 76.8\% \\ \cline{3-5}
 & & all & 5, 200, 0.001, 0.3, $10^4$, 45 & 77.8\% \\ \hline
\end{tabular}
\caption{The selected networks for $M_\rho=5$ TeV. $N_{\rm epoch}$ is the epoch number when we cut the training.}
\label{tab:best_network_5}
\end{table*}

The selected DNN configurations for $M_\rho=1$ and 5 TeV are listed in Table~\ref{tab:best_network_1} and Table~\ref{tab:best_network_5}, respectively. The selection criteria are described in Section~\ref{sec:training}. The epochs when we cut the training are listed in the forth columns. One can see that for a individual signal benchmark in a given kinematic region, the DNN with low-level observables usually requires a longer training epoch than the DNN with all observables, if they have the same configurations. That is because the DNN needs more time to learn about the physics in the signal process, if no hint is given to it. The classification accuracies (on the validation/test data) of the networks are given in the fifth columns.

\begin{table*}
\scriptsize\renewcommand\arraystretch{1.5}\centering
\begin{tabular}{|c|c|c|c|c|c|c|c|c|c|c|c|c|c|c|c|c|c|}\hline
Models & \multicolumn{2}{|c|}{M1$\Gamma$1} & \multicolumn{2}{|c|}{M1$\Gamma$2} & \multicolumn{2}{|c|}{M1$\Gamma$3} & \multicolumn{2}{|c|}{M1$\Gamma$4} & M5$\Gamma$1 & M5$\Gamma$2 & M5$\Gamma$3 & M5$\Gamma$4 \\ \hline
Kinematic region & resolved & boosted & resolved & boosted & resolved & boosted & resolved & boosted & boosted & boosted & boosted & boosted \\ \hline
Low-level input & 85.2\% & 67.9\% & 83.2\% & 65.8\% & 81.6\% & 65.1\% & 80.8\% & 64.3\% & 79.5\% & 78.2\% & 77.4\% & 76.8\%\\ \hline
Planing away $M_{t\bar t}$ & 76.8\% & 63.7\% & 75.3\% & 62.7\% & 74.1\% & 62.1\% & 73.0\% & 62.3\% & 65.3\% & 65.8\% & 63.7\% & 64.1\% \\ \hline
\end{tabular}
\caption{The accuracy reach of the chosen neural networks before and after planing away $M_{t\bar t}$. The configurations of the DNN's are listed in Tables~\ref{tab:best_network_1} and \ref{tab:best_network_5}.}
\label{tab:planing}
\end{table*}

Table~\ref{tab:planing} shows the accuracy reach of the DNNs before and after planing away the key observable $M_{t\bar t}$. The data of the second row, i.e. the accuracies before planing, are taken from the fifth columns of Tables~\ref{tab:best_network_1} and \ref{tab:best_network_5}. While the accuracies after planing listed in the third row are obtained by the weighted training described in Section~\ref{sec:planing}.

\bibliographystyle{apsrev}
\bibliography{reference}

\end{document}